\documentclass[11pt,3p,review,authoryear]{elsarticle}

\usepackage{amsmath,amssymb,amsthm,mathtools,bm}

\usepackage{graphicx}
\usepackage{epstopdf}
\usepackage{subcaption}
\usepackage{csvsimple}
\usepackage{tabularx}
\usepackage{longtable}
\usepackage{booktabs}
\usepackage{pdflscape}
\usepackage{ltablex}
\usepackage{enumitem}
\usepackage{siunitx}
\usepackage{multirow}
\usepackage{threeparttable}
\usepackage{tikz}
\usetikzlibrary{arrows.meta,positioning,shapes.geometric}

\usepackage[hidelinks]{hyperref}
\usepackage{cleveref}

\journal{International Journal of Forecasting}
\biboptions{longnamesfirst}

\begin{document}

\begin{frontmatter}

\title{\texorpdfstring{Explainable Machine Learning for Macroeconomic and Financial Nowcasting:\\A Decision‑Grade Framework\\for Business and Policy}
{Explainable Machine Learning for Macroeconomic and Financial Nowcasting}}

\author[unimc,essec]{Luca Attolico\corref{cor}}

\address[unimc]{Dipartimento di Economia e Diritto, Università degli Studi di Macerata, Italy}
\address[essec]{ESSEC Business School, Asia-Pacific, Singapore}

\cortext[cor]{Corresponding author}
\ead{luca.attolico@unimc.it}
\ead[url]{https://orcid.org/0000-0002-8649-5440}

\begin{abstract}
Macroeconomic nowcasting sits at the intersection of traditional econometrics, data‑rich information systems, and AI applications in business, economics, and policy. Machine learning (ML) methods are increasingly used to nowcast quarterly GDP growth, but adoption in high‑stakes settings requires that predictive accuracy be matched by interpretability and robust uncertainty quantification. This article reviews recent developments in macroeconomic nowcasting and compares econometric benchmarks with ML approaches in data‑rich and shock‑prone environments, emphasizing the use of nowcasts as decision inputs rather than as mere error‑minimization exercises. The discussion is organized along three axes. First, we contrast penalized regressions, dimension‑reduction techniques, tree ensembles, and neural networks with autoregressive models, Dynamic Factor Models, and Random Walks, emphasizing how each family handles small samples, collinearity, mixed frequencies, and regime shifts. Second, we examine explainability tools—intrinsic measures and model‑agnostic XAI methods—focusing on temporal stability, sign coherence, and their ability to sustain credible economic narratives and nowcast revisions. Third, we analyze non‑parametric uncertainty quantification via block bootstrapping for predictive intervals and confidence bands on feature importance under serial dependence and ragged edge. We translate these elements into a reference workflow for ``decision‑grade'' nowcasting systems, including vintage management, time‑aware validation, and automated reliability audits, and we outline a research agenda on regime‑dependent model comparison, bootstrap design for latent components, and temporal stability of explanations. Explainable ML and uncertainty quantification emerge as structural components of a responsible forecasting pipeline, not optional refinements.
\end{abstract}

\begin{keyword}
macroeconomic and financial nowcasting \sep explainable machine learning \sep uncertainty quantification \sep structural breaks \sep policy applications
\end{keyword}

\end{frontmatter}

\section[Introduction]{Introduction}
\label{sec:intro}

Timely and reliable macroeconomic indicators are a prerequisite for sound economic policy and portfolio management. Yet official statistics such as GDP growth, inflation, and employment are published with substantial delays relative to the reference period. This creates an ``information gap'' \citep{GiannoneReichlinSmall2008,BokEtAl2018} that forces policy makers and private agents to act under high uncertainty. Nowcasting addresses this gap: instead of projecting the distant future, it delivers estimates consistent with current conditions by exploiting available high‑frequency indicators \citep{BaumeisterEtAl2021,BanburaRunstler2011}, thereby reducing information latency and improving the timeliness of fiscal, monetary, and market responses.

This article reviews the state of the art in macroeconomic nowcasting with a specific focus on machine learning (ML) contributions in data‑rich environments. We compare these methods with traditional econometric approaches, taking explainability and uncertainty quantification as core requirements for decision‑grade use. The scope is real‑activity nowcasting (with a constant reference to GDP growth) in heterogeneous and high‑dimensional datasets, and environments characterized by shocks and regime changes that stress linear and stationarity assumptions. The analysis explicitly extends to the macro‑financial dimension, allowing for market variables and credit conditions (e.g., interest rates, spreads, volatility) as anticipatory drivers of GDP dynamics.

The stakeholder base is broad. Central banks and finance ministries share the decision space with commercial banks, insurers, asset managers, corporate research units, and rating agencies, which use nowcasts for risk management, tactical asset allocation, budgeting, and investor communication. This diversity increases the demand for real‑time signals, robust uncertainty metrics, and clear economic narratives underpinning the nowcast.

The COVID‑19 shock made these requirements explicit. In 2020, standard release calendars failed to provide sufficiently prompt signals of turning points. High‑frequency indicators and mobility measures, combined with institutional tools such as GDPNow (Atlanta Fed) and the Staff Nowcast (NY Fed) \citep{Higgins2014,BokEtAl2018,CajnerEtAl2020,ForoniMarcellinoStevanovic2020}, allowed near‑real‑time tracking of shock intensity and recovery.

The economic rationale is twofold. First, nowcasting reduces information asymmetry between the state of the economy and the instruments available to policy makers and market participants, shortening the reaction horizon and improving the effectiveness of discretionary or rule‑based measures. Second, it allows systematic scenario updates in the presence of non‑linear shocks (regime changes, structural breaks, reopenings) where linear models are misspecified \citep{MedeirosEtAl2021,CoulombeEtAl2022}. In modern macro‑financial datasets, which are large, noisy, and strongly correlated, combinations of regularization, dimension reduction, and ensembling typical of ML increase the ability to extract signals and adapt to evolving dynamics.

Our framework rests on three technical premises:
\begin{enumerate}[label=\roman*), leftmargin=*, itemsep=0pt]
    \item useful information is distributed across many series, with pervasive correlations and multicollinearity;
    \item non‑linearities and interactions are particularly relevant around shocks and recoveries;
    \item adoption in policy‑making contexts requires explainability tools for advanced regression methods and robust predictive uncertainty quantification that respect serial dependence and potential breaks.
\end{enumerate}

Literature and practice converge on a single tenet: accuracy without transparency and uncertainty measures is insufficient \citep{Rudin2019,PetropoulosMakridakis2020} for policy purposes and financial accountability. Explainability is not an auxiliary property but a structural requirement for analytics governance: it must clarify which variable families drive the nowcast, how their roles shift across regimes, and with what statistical confidence. Our perspective explicitly integrates feature‑importance explainability with predictive interval estimation \citep{LundbergEtAl2020,SundararajanTalyYan2017,PolitisRomano1994,Li2021}, prioritizing procedures consistent with the temporal dependence of the data.

The review is organized around methods, trade‑offs, and operational implications. The focus remains on macroeconomics and macro‑finance: translating high‑frequency signals into current‑activity measures that can be used in institutional and business settings. A well‑designed nowcasting framework should improve turning‑point detection, reduce timing errors in policy responses, support more disciplined tactical portfolio adjustments, and strengthen external communication by aligning explanations with the model structure. The combination of methodological comparison (traditional econometrics vs.\ ML), feature‑importance explainability, and predictive uncertainty quantification provides a coherent set of ``decision aids'' that mitigates misspecification and overfitting risks, increasing the credibility of nowcast‑based messaging.

This article contributes to the nowcasting and forecasting literature in three ways. First, it jointly reviews econometric benchmarks and modern ML methods through the lens of decision‑grade use, tying model performance to explainability and uncertainty rather than to accuracy alone. Second, it systematizes the role of feature‑importance explainability and non‑parametric, block bootstrap based predictive intervals in macro‑financial nowcasting, emphasizing their implications for governance and communication. Third, it translates these requirements into a conceptual end‑to‑end workflow for decision‑grade nowcasting and into a research agenda explicitly oriented towards institutional and market deployment.

The remainder of the paper is organized as follows. Sections~\ref{sec:catalyst} and~\ref{sec:sota} discuss the main nowcasting model families, from econometric benchmarks to ML methods. Section~\ref{sec:prereq} analyzes feature importance explainability tools and their application over rolling windows. Sections~\ref{sec:implications} and~\ref{sec:workflow} derive operational implications for business, economics, and policy, and outline a conceptual workflow for decision‑grade nowcasting systems. Section~\ref{sec:agenda} discusses the future research agenda, with particular attention to non‑parametric uncertainty quantification.

\section[Machine learning as a catalyst for nowcasting]{Machine learning as a catalyst for nowcasting}
\label{sec:catalyst}

Machine learning applications in macroeconomic and macro-financial nowcasting occupy the intersection of traditional econometrics and data-rich information systems. In business, economic, and policy environments characterized by dense information flows (spanning diverse frequencies, administrative series, market variables, and high-frequency measures), classical linear models remain a benchmark but often prove insufficient. Machine learning acts as a catalyst \citep{MedeirosEtAl2021,CoulombeEtAl2022,KimSwanson2018}: it does not replace the econometric framework but integrates a toolkit capable of handling high dimensionality, collinearity, non-linearity, and regime breaks without imposing a rigid ex-ante functional specification. Under the classical assumption of independent and identically distributed (i.i.d.) data, the robustness of many learning algorithms is supported by learning theory,\footnote{See, for example, \citet{mohri2018foundations} and \citet{bach2024learning} for generalization bounds and consistency conditions.} which provides formal generalization guarantees.

In nowcasting problems, short time samples, large predictor sets, mixed frequencies, and publication lags are the norm. The suite of methods considered in this review, comprising penalized regressions (LASSO, Ridge, Elastic Net), dimension reduction techniques (Principal Component Regression, Partial Least Squares Regression), tree ensembles (Random Forest, Gradient Boosted Decision Trees), and neural networks (Multilayer Perceptrons, Gated Recurrent Units), covers four complementary strategies to address these constraints: complexity reduction via regularization, information compression into latent factors, non-parametric modeling of interactions, and explicit learning of sequential dynamics. In all cases, we employ linear econometric benchmarks (Random Walk, Autoregressive models, and Dynamic Factor Models) \citep{ZouHastie2005,JolliffeCadima2016,Wold1985,Breiman2001,ChenGuestrin2016,ChoEtAl2014,HewamalageBergmeirBandara2021,StockWatson2002,BaiNg2008,BanburaEtAl2010}, which remain the reference standard for stability and interpretation, but are augmented by more flexible tools when data structure demands it.

Penalized regressions constitute the first block. In the presence of numerous correlated regressors and very short forecast horizons, LASSO, Ridge, and Elastic Net---employing L1 (sparsity), L2 (shrinkage), and flexible L1/L2 penalties respectively---offer a transparent trade-off between parsimony and information utilization. LASSO selects variable subsets by imposing sparsity; Ridge retains all variables but stabilizes coefficients; Elastic Net is beneficial when variables are organized in highly correlated groups \citep{ZouHastie2005,SmeekesWijler2018,UematsuTanaka2019} and pure L1 penalties become unstable. Analytical validation is provided by \citet{alquier2011sparsity}, who show---under specific temporal dependence and mixing hypotheses---that sparsity properties and oracle inequalities for penalized regressions extend to time-dependent contexts, providing a formal basis for their use in macroeconomic series. More recently, \citet{wong2020lasso} demonstrated that, under $\beta$-mixing conditions and sufficiently strong moments, LASSO retains selection and prediction guarantees even for heavy-tailed series, a frequent case in macro-financial data. The advantage is twofold: it stabilizes inference in high dimensions and preserves a ``coefficient-based'' structure that facilitates economic interpretation of drivers, particularly within an ex-post explainability framework. Given their intrinsically linear structure, however, these models cannot autonomously capture non-linear relationships; such features must be explicitly constructed ex ante and incorporated into the dataset.

A second block consists of Principal Component Regression (PCR) and Partial Least Squares Regression (PLSR), which compress large sets of predictors into a few components. PCR constructs factors that maximize the explained variance among predictors; PLSR orients components towards prediction by maximizing the covariance between predictors and the target. With limited time samples and high-dimensional macroeconomic datasets, a few components can summarize common dynamics, reducing noise and coefficient instability. Operationally, the impact on nowcasting is often greater out-of-sample stability compared to complete regressions in the presence of near-perfect multicollinearity, albeit at the cost of reduced transparency: components are linear combinations of many variables, and their ``economic reading'' requires dedicated tools, such as Variable Importance in Projection (VIP) for PLSR, to map the relevance of original covariates back onto the latent factors \citep{JolliffeCadima2016,MehmoodEtAl2012,BoivinNg2006,Wold1985}.

Ensemble learning models, specifically Random Forest (RF) and Gradient Boosted Decision Trees (GBDT), address the problem non-parametrically by constructing adaptive partitions of the predictor space to model non-linearities and interactions. RF relies on bagging, that is, averaging many trees trained on bootstrap samples
and random feature subsets; GBDT implements boosting with explicit regularization, iteratively focusing on the residuals to increase predictive capacity. Recent empirical evidence on episodes of high macroeconomic instability, however, suggests that without stringent discipline (e.g., on tree depth, minimum samples per leaf, learning rate, and temporal validation), ensemble models risk overfitting rare regimes and degrading performance during shocks \citep{ProbstWrightBoulesteix2019,MedeirosEtAl2021,CoulombeEtAl2022}. Theoretical support for this architecture is provided by \citet{goehry2020random}, who extends consistency results for Random Forests to stationary time-dependent processes, showing that these architectures can learn reliably even in the presence of serial dependence.

Neural networks, such as Multilayer Perceptron (MLP) and Gated Recurrent Unit (GRU), further enhance flexibility. An MLP serves as a universal approximator for non-linear functions of static predictors; a GRU introduces a latent state that evolves, making it natural to model sequential dependencies and memory effects \citep{ChoEtAl2014,HewamalageBergmeirBandara2021,GoodfellowBengioCourville2016,Tashman2000,HyndmanAthanasopoulos2021}. In macroeconomic and macro-financial nowcasting, these models are valuable when integrating high-frequency flows or when the relationship between leading indicators and quarterly growth exhibits local discontinuities, thresholds, or complex interactions. Their representational capacity, however, is contingent on controlling overfitting risk on limited-time samples, the quality of tuning (layer size, dropout, learning rate, backpropagation horizon), and the design of validation procedures in a time-series environment. Without rigorous experimental discipline, gains in accuracy in ``normal'' conditions may be nullified by marked instability during crisis regimes. Recent advances in learning theory for deep networks on weakly dependent processes \citep{kengne2024deep} clarify the conditions under which neural architectures can guarantee generalization capabilities even in the presence of temporal dependence.

A cross-cutting issue for all model families is their suitability for decision-grade use in business and policy contexts. Linear regularized methods and component-based techniques offer more immediate explanatory avenues, as their structure remains close to traditional econometric models; conversely, ensembles and networks explicitly require explainability and calibrated uncertainty measures to be employed responsibly \citep{Rudin2019,LundbergEtAl2020,SundararajanTalyYan2017,PolitisRomano1994,Li2021}. In a macroeconomic environment with high variable counts, publication lags, and potential regime breaks, no single model class is dominant across all scenarios.

\section[State of the art: from econometrics to machine learning]{State of the art: from econometrics to machine learning}
\label{sec:sota}

The evolution of macroeconomic nowcasting has followed two primary strands: traditional econometric approaches, with Dynamic Factor Models (DFMs) as the cornerstone \citep{StockWatson2002,StockWatson2011,BaiNg2002}, and a more recent lineage based on ML methodologies. The literature credits DFMs with the ability to synthesize common information from large datasets into a few latent factors, thereby improving nowcasting efficiency when many series are correlated and redundant. In modern state-space implementations, the Expectation-Maximization (EM) algorithm and the Kalman filter allow for systematic handling of ragged edge, missing data, and mixed frequencies, expanding their utility in institutional contexts \citep{DozGiannoneReichlin2011,DozGiannoneReichlin2012,BanburaRunstler2011}. The main vulnerability of DFMs lies in the choice and stability of the factor space: determining the number of factors and their economic interpretation relies on partly arbitrary statistical criteria, which can lead to misspecification—specifically, an incomplete representation of common dynamics—particularly in large-scale datasets where a few stable factors do not fully capture the common component \citep{BaiNg2002,BanburaEtAl2010}. A structural constraint also applies: the assumption of predominantly linear relationships between factors and observed variables \citep{StockWatson2011}. This hypothesis reduces the capacity to describe non-linearities and interactions that tend to emerge precisely during phases of instability (e.g., financial crises, pandemics, energy shocks).

Autoregressive approaches (AR and VAR) constitute an established alternative, favored for their simplicity, transparency, and immediately readable statistical grammar \citep{Hamilton1994}. AR and VAR models perform well in small dimensions and relatively stable regimes, offering traceable coefficients and impulse response functions. Alongside these, the Random Walk (RW), with or without a deterministic drift term, serves as a minimal, transparent baseline---often competitive with highly persistent series and valuable as a benchmark to evaluate the actual gains of richer models \citep{NelsonPlosser1982,StockWatson1996,MarcellinoStockWatson2006}. Their vulnerability lies in the ``curse of dimensionality'': including many regressors and lags rapidly leads to instability and overfitting. Where DFMs compress information into factors, AR and VAR models require extreme parsimony or selection; otherwise, out-of-sample quality degrades, especially in the presence of structural breaks.

This dual landscape, with DFMs as a ``globally parsimonious'' linear synthesis and AR/VAR as transparent but high-dimensional, fragile benchmarks, has driven the adoption of tools capable of handling rich datasets, non-linear interactions, and shifting regimes without imposing structure on a few factors or rigid parametric schemes. The literature documents the entry of ML into macro nowcasting with empirical motivations: superior predictive properties when the signal-to-noise relationship is complex and dynamics change irregularly; greater flexibility in managing heterogeneity, collinearity, and mixed-frequency signals; and regularization and automatic selection mechanisms to discipline high dimensionality \citep{Varian2014,MedeirosEtAl2021,CoulombeEtAl2022,KimSwanson2018}. Comparative evidence, reconstructed in reviews and applied studies, attributes superior robustness to ML in non-linear contexts, provided that temporal validation procedures and complexity controls are in place to avoid out-of-sample instability due to changes in the data generating process over time \citep{Tashman2000,InoueRossi2012,HyndmanAthanasopoulos2021,PesaranTimmermann2007,ClarkMcCracken2009}. Methodological work has extended generalization guarantees from the i.i.d. case to dependent processes, including non-parametric regression and SVM-type algorithms, under various mixing conditions \citep{meir2000nonparametric,steinwart2009learning,mohri10a}.

The analytical perspective of DFMs remains competitive when the common component is dominant and reasonably stable, ragged edge is a central issue, and communication requires compact factor-based summaries. Under these conditions, a few factors explain a large portion of the covariance, Kalman filtering reduces information latency in the presence of staggered releases \citep{BanburaRunstler2011,DozGiannoneReichlin2011,DozGiannoneReichlin2012}, and the linear structure offers clear economic levers: factors can be read as syntheses of information blocks (cycles, industry, prices), and coefficients retain a stable meaning.

ML methods add value when useful signal concentrates in non-linear combinations (thresholds, interactions, regime-specific effects), when frequent regime changes weaken the stability of factor loadings, or when high-frequency indicators are central and the ragged-edge advantage of DFMs is limited because series are already harmonized. In these environments, ensembles and neural networks can capture residual signals that linear factors do not represent well, but at the cost of an “explainability debt”: feature importance, local attributions, and sensitivity analyses must be deployed to restore economic traceability and make ML-based nowcasts admissible for decision-grade use.

Reported comparisons indicate recurrent patterns. In phases of relative macroeconomic stability, well-calibrated DFMs compete closely with machine learning solutions and linear benchmarks (autoregressive models and simple RWs), especially when the common component is marked and covariates are numerous but largely redundant. In turbulent phases, models capable of capturing interactions and non-linearities (e.g., boosted trees or memory networks) tend to retain proper predictive signals longer, provided complexity is disciplined (e.g., tree depth, regularization, early stopping) and validation is conducted with walk-forward schemes. However, the advantage is not guaranteed: with short samples, persistent shocks, and high noise, even the most flexible ML specifications can lose much of their margin over simpler models, yielding results comparable to autoregressive models or a simple RW. The robustness of models, whether factorial, autoregressive, or machine learning, depends mainly on the experimental design and on the respect for the data's temporal structure.

Regarding high-dimensionality management, the pathways are distinct. DFMs compress information through a few factors, effectively imposing a ``low-dimensional'' structure on the predictor space. In ML, reduction occurs via regularization and selection, with potential further levels of latent representation in neural networks. The result is comparable in terms of objective (avoiding variance and overfitting), but different in implications: factors have an aggregate semantics, whereas ML solutions maintain a more direct link to the original features (without latent layers), allowing for specific-variable predictive-contribution analyses. That explains why studies emphasizing diagnosis and macro communication often prefer DFMs, while contexts of high-frequency surveillance and digital signals favor ML specifications with supporting attribution tools.

Resilience to shocks and breaks deserves distinct attention. The linearity of DFMs, ARs, and VARs can prove too restrictive when shocks rapidly modify relationships and variances. Factor estimation updated in real-time mitigates the problem but does not eliminate it; loadings and the latent dimensionality may not change gradually. In ML, adaptation occurs through explicit strategies: frequent updates; stringent regularization;
control of changes in the data-generating process over time and of parameter instability; validation that mirrors the temporal sequence. On the theoretical front, recent contributions also develop generalization bounds for non-stationary and non-mixing sequences, based on notions of discrepancy between distributions over time \citep{kuznetsov2020discrepancy}, which are particularly relevant in the presence of persistent regime changes. Literature exploring episodes of strong instability signals a relative advantage for well-regularized ML, consistent with the idea that accuracy in the presence of non-linearities and interactions matters most precisely when the economic structure shifts \citep{GiannoneLenzaPrimiceri2021,LenzaPrimiceri2022,PesaranTimmermann2007,ClarkMcCracken2009}.

Table~\ref{tab:ml_nowcasting_lit} summarizes a set of representative empirical applications of machine learning to macroeconomic and macro-financial nowcasting. For each study, it reports the target variable, sample, main model families, and performance relative to econometric benchmarks, thereby highlighting the heterogeneity of regimes and design choices.

\begin{footnotesize}
\begingroup
\setlength{\tabcolsep}{3pt}
\renewcommand{\arraystretch}{1.05}
\begin{longtable}{%
  >{\raggedright\arraybackslash}p{0.17\textwidth}%
  >{\raggedright\arraybackslash}p{0.24\textwidth}%
  >{\raggedright\arraybackslash}p{0.23\textwidth}%
  >{\raggedright\arraybackslash}p{0.24\textwidth}%
}
\caption{Selected empirical applications of data-rich and machine-learning methods to macroeconomic nowcasting and short-horizon forecasting.}
\label{tab:ml_nowcasting_lit}\\
\toprule
Study & Context and target & Methods & Main findings (vs benchmarks) \\
\midrule
\endfirsthead
\toprule
Study & Context and target & Methods & Main findings (vs benchmarks) \\
\midrule
\endhead
Bańbura and Rünstler (2011) &
Euro area GDP and components; role of hard vs soft indicators under publication lags. &
Large-scale DFM in state-space form; EM--Kalman estimation for mixed frequencies and ragged edge. &
DFM exploits surveys and hard data to deliver timely, accurate nowcasts, outperforming small AR benchmarks. \\[0.6ex] \midrule

Bok, Caratelli, Giannone, Sbordone and Tambalotti (2018) &
US and euro-area real-activity nowcasting with large macro-financial panels in true real time. &
Data-rich DFMs and Bayesian VARs tailored to institutional nowcasting (e.g., NY Fed Staff Nowcast). &
Big-data DFMs materially improve nowcasting performance, especially around turning points, versus simple AR and small VAR benchmarks. \\[0.6ex] \midrule

Baumeister, Korobilis and Lee (2021) &
Global activity and energy markets; link between oil prices and global real conditions. &
Factor and Bayesian multivariate models extracting common components from macro and energy variables. &
Energy-market factors are powerful drivers of global activity and yield forecasts competitive with parsimonious benchmarks. \\[0.6ex] \midrule

Kim and Swanson (2018) &
US macro-financial forecasting in a big-data setting with very large predictor sets. &
Factor models, forecast combinations, and shrinkage methods (ridge, LASSO) applied to large predictor panels. &
Combining factors with shrinkage and forecast pooling yields robust gains over small-scale models when many predictors are available. \\[0.6ex] \midrule

Medeiros, Vasconcelos, Veiga and Zilberman (2021) &
Short-horizon US inflation forecasting in a data-rich macro-financial environment. &
Penalized regressions and ML models (LASSO, tree-based methods, neural networks) benchmarked against DFMs and linear models. &
Regularized ML often outperforms linear benchmarks in nonlinear and unstable regimes, provided validation is strictly time-aware. \\[0.6ex] \midrule

Coulombe, Leroux, Stevanovič and Surprenant (2022) &
US output, inflation and labor-market forecasting across distinct macro regimes. &
Portfolio of ML methods (regularized linear models, tree ensembles, neural networks) versus standard econometric benchmarks. &
ML gains concentrate in unstable periods and for selected targets; in stable regimes simple models can match or dominate complex specifications. \\
\bottomrule
\end{longtable}
\endgroup
\end{footnotesize}

These outcomes do not reduce econometric approaches to an ancillary role but point to complementarity. Solid DFMs, in the presence of a clear common component, provide parsimonious and explainable baselines, useful also for benchmarking and methodological stress testing. ML methods add value when the relational structure is more complex, provided one controls for leakage, importance distortions under correlated features, and employs dedicated tools to restore economic traceability. In both cases, a validation and reporting framework that includes explicit explainability and uncertainty quantification is the enabling condition for nowcasting results to be treated as decision-grade and to maintain credibility over time.

\section[Methodological prerequisites: feature importance explainability and non-parametric uncertainty quantification]{Methodological prerequisites: feature importance explainability and non-\\parametric uncertainty quantification}
\label{sec:prereq}

Business and economics are high-stakes domains in which public and private decisions are tracked, contestable, and subject to ex post scrutiny. A nowcast devoid of explanations and uncertainty measures fails to meet minimum transparency and accountability requirements. Rigorous explainability and statistical variability quantification are not optional add-ons, but prerequisites for the operational deployment of ML models in macroeconomic environments \citep{Rudin2019}. The objective is twofold: to explicate the economic drivers underpinning the nowcast and to provide predictive intervals that reflect, with temporal coherence, the plausible dispersion around the point forecast.

We define ``explainability'' as the ability to link nowcast variations to feature changes in a stable and controllable manner, distinguishing between the local level (explanations for a single vintage) and the global level (average patterns over time). XAI tools are categorized into model-specific (leveraging the model's internal structure) and model-agnostic (applicable across heterogeneous classes). Both dimensions are necessary in nowcasting: local, because decision-makers operate on real-time releases; global, to ensure narrative consistency regarding the role of macro drivers across different regimes. Three practical properties guide their adoption in economic time series: \textbf{stability} (robustness to perturbations and initialization), \textbf{temporal coherence} (absence of arbitrary oscillations between vintages given identical information), and \textbf{economic compatibility} (absence of signals contradicting fundamental macro identities).

\textit{Integrated Gradients} (IG) is a model-specific tool for neural networks (MLP, GRU) that assigns a local contribution to each variable by integrating the model's gradients along a path from a baseline to the input. The axiom of completeness makes the accounting of contributions particularly clear. The baseline must be defined ex-ante: a zero vector is common but not always economically or financially interpretable; alternative baselines (e.g., pre-shock historical mean or calibration-window median) allow for sensitivity analysis of attributions and robustness checks. In real-time applications, using IG on GRUs enables attribution across the entire information sequence, clarifying whether a nowcast revision stems from new data arrivals or from the reweighting of existing signals. Operationally, IG attributions are locally signed relative to the baseline; globally, one may summarize their magnitude (e.g., via the mean of absolute values) or preserve the sign when relevant. Reporting should explicitly distinguish between these two readings to avoid interpretive ambiguity \citep{SundararajanTalyYan2017}.

\textit{SHapley Additive exPlanations} (SHAP) is a model-agnostic method that locally decomposes the output into a base term and a set of feature contributions, which is useful for auditing and vintage comparison. Two caveats are central. With highly correlated features, credit assignment can become conventional: the metric remains internally consistent, but identification of a single driver is fragile. Moreover, one typically relies on approximations,\footnote{An exception is the tree-ensemble family, for which the TreeExplainer algorithm \citep{LundbergEtAl2020} computes exact Shapley values in polynomial time.} so explanations should be computed out-of-sample using samplers that preserve temporal dependence; time-aware kernels and background sets built on contiguous blocks help respect chronological order and prevent leakage. Pairing ``global'' SHAP (average patterns) with ``local'' IG (instantaneous mechanics) facilitates cross-checking over rolling windows.

\textit{Permutation feature importance}, also model-agnostic, quantifies accuracy loss when the link between a variable and the output is broken \citep{Breiman2001}. In nowcasting, permutations must respect temporal and cross-sectional dependence, ideally via block or conditional variants that avoid destroying the joint structure of the predictors. Its most robust application is as a stress test: null degradation indicates weak dependence, whereas significant and unstable degradations across vintages signal local overfitting risks.

Useful explainability requires a governance pipeline, not merely the existence of a metric. In practice, this involves calculating out-of-sample attributions and measuring their stability over rolling windows. It also requires aligning scale and sign with macro conventions (e.g., positive contributions of consumption and investment to GDP growth), flagging areas of interpretive uncertainty (quasi-collinear clusters), and documenting pre-processing rules that affect semantics (normalizations, log-transforms). Audit materials should include at least: a global map of drivers, local explanations for the latest vintages, and sensitivity of attributions to reasonable choices of baseline and sampling.

Real-time uncertainty quantification requires procedures consistent with serial dependence and release asynchrony. Block bootstrapping (e.g., moving-block or stationary bootstrap) is a natural solution, under assumptions of sufficiently regular serial dependence, when avoiding strong parametric assumptions is desired: contiguous sequences of observations (blocks) are sampled to generate alternative input trajectories, and the nowcast is recalculated along each trajectory \citep{PolitisRomano1994,Li2021}. That yields an empirical distribution of the nowcast from which prediction intervals are derived. Block length must be calibrated to preserve the autocorrelation of the target and principal predictors: blocks that are too short break dynamics, whereas blocks that are too long reduce the number of effective trajectories. In practice, one checks how coverage and interval width behave over neighboring lengths and selects a range that yields stable results.

Table~\ref{tab:xai_uncertainty_framework} summarizes how the proposed framework allocates feature importance explainability and uncertainty quantification across the main model families used for nowcasting. This structured view clarifies the role of each family in decision-grade applications and underscores the current lack of standardized, cross-family practices.

\begin{footnotesize}
\begingroup
\setlength{\tabcolsep}{3pt}
\renewcommand{\arraystretch}{1.05}

\begin{longtable}{%
  >{\raggedright\arraybackslash}p{0.18\textwidth}%
  >{\raggedright\arraybackslash}p{0.26\textwidth}%
  >{\raggedright\arraybackslash}p{0.26\textwidth}%
  >{\raggedright\arraybackslash}p{0.26\textwidth}%
  }
\caption{Explainability and uncertainty quantification by model family in the proposed nowcasting framework.}
\label{tab:xai_uncertainty_framework}\\
\toprule
Model family & Explainability tools & Uncertainty quantification & Notes for decision-grade nowcasting \\
\midrule
\endfirsthead

\toprule
Model family & Explainability tools & Uncertainty quantification & Notes for decision-grade nowcasting \\
\midrule
\endhead

Penalized regressions (LASSO, Ridge, Elastic Net) &
Standardized coefficients; regularization paths; rolling-window stability checks. Block-bootstrap confidence bands by economic block (domestic demand, external sector, financial conditions). Conditional, time-aware permutation importance as incremental-contribution stress test. &
Time-series block bootstrap on the full predictor-target system with re-estimation on each resample. Empirical forecast distribution for prediction intervals; same draws reused for bands on coefficient-based importance. &
Linear, structurally transparent baseline for data-rich setups. Handles short samples and collinearity; coefficients and bands support readable macro narratives and reliability of identified drivers. \\[0.4ex]
\midrule

Dimension‑\linebreak[1]reduction models (PCR, PLSR) &
Component loadings and component contributions to the target. VIP scores (PLSR) to map predictors back to components. Component-level permutation importance, with attention to correlated factors and temporal dependence. &
Block bootstrap on original series with recomputation of components, yielding empirical intervals for forecasts and component contributions. &
Compact representation of large panels, with explanations organized around economically interpretable components. Bootstrap intervals quantify stability of the latent structure and robustness of component-level importance across regimes. \\[0.4ex]
\midrule

Tree ensembles (Random Forest, Gradient Boosted Trees) &
Global and local SHAP values (TreeExplainer) with time-aware background sets. Split-based importances used only as secondary diagnostics. Block or conditional permutation importance to respect serial and cross-sectional dependence; SHAP-profile stability monitored over rolling windows. &
Block bootstrap on time-indexed observation or residual blocks, with trees re-fitted or updated on each resample. RF tree bootstrap combined with time-aware resampling of the training set. Empirical distributions used for prediction intervals and uncertainty bands around SHAP-based importance. &
Captures nonlinearities and interactions without explicit functional form, but with significant explainability debt. SHAP + block/conditional permutation + bootstrap bands make ensemble-based nowcasts auditable and limit overinterpretation of unstable importance rankings. \\[0.4ex]
\midrule

Neural networks (MLP, GRU) &
Integrated Gradients (IG) with economically meaningful baselines (pre-shock mean or calibration-window median) and path-averaging. For GRUs, IG over full sequences to separate effects of new releases from reweighting of past information. Gradient-based and SHAP-style variants as cross-checks; time-aware permutation importance as stress test. &
Block bootstrap on sequences of time-indexed inputs consistent with the release calendar, with re-estimation or fine-tuning on each resample. Empirical forecast distributions for prediction intervals; IG recomputed on bootstrap models to obtain attribution bands. Diagnostics on coverage, interval asymmetry near turning points, and stability of IG profiles. &
Maximum flexibility for nonlinear, regime-dependent dynamics, but higher model risk. IG-based explanations plus bootstrap intervals on forecasts and attributions are required for neural nowcasts to be admissible in high-stakes policy or market use. \\
\bottomrule
\end{longtable}
\endgroup
\end{footnotesize}

With heterogeneous series and ragged edge, blocks must be constructed on ``observed'' vectors consistent with the publication calendar: frequency harmonization and imputation techniques must respect realistic availability patterns, avoiding spurious scenarios that include unreleased variables. If the model has latent components or state filters, bootstrapping can be applied to measurement residuals or state innovations; in purely reduced-form ML models, the most transparent approach remains block resampling of predictors, with a complete pipeline re-fit on each sample, thereby incorporating parameter estimation uncertainty.

Prediction intervals require accurate diagnostics. Nominal coverage and average length constitute a first check. However, dynamic consistency is central in macroeconomic contexts: widths should expand near turning points, narrow when the common informative component is more pronounced, and show plausible asymmetries during structural breaks. An excessively flat profile does not signal stability, but rather inadequate calibration of the bootstrap procedure. Sequential comparison between iterations of the \textit{walk-forward} nowcasting scheme also provides an indicator of communicative stability: large recalibrations unjustified by new information highlight fragility in the experimental design or estimation pipeline \citep{Tashman2000,HyndmanAthanasopoulos2021}.

Explainability and uncertainty remain methodologically distinct planes, but must be read jointly. Volatile attributions render the origin of nowcast variations opaque; intervals that are too narrow or too wide erode the credibility of the attributions themselves. A pragmatic integration holds the two aspects together with three coordinated moves, without confusing levels:
\begin{enumerate}[label=\roman*), leftmargin=*, itemsep=0pt]
    \item mapping the contributions of effectively active variables, combining local readings (relative to the single quarterly prediction) and global readings (relative to the entire nowcasting period and its sub-periods), to distinguish episodic contributions from persistent factors;
    \item adopting block bootstrapping so that predictive intervals are calibrated to respect serial dependence and potential heterogeneity across business cycle phases;
    \item producing a joint report of feature importance measures and intervals that allows expert readers to evaluate the intensity and origin of nowcast variations simultaneously.
\end{enumerate}

Analytical responsibility translates into conservative choices regarding pre-processing and importance measures. Intensive feature engineering increases the risk of collinearity and artifactual attributions; a compact and readable predictor set, modified only for motivated economic reasons, favors explanation stability. Importances must be accompanied by uncertainties (intervals via resampling) and sensitivity controls, including subsampling, exclusion of groups with high information overlap, and comparisons across multiple attribution methods to avoid tool-dependent conclusions. Monotonic transformations consistent with theory (logs on levels, differences on rates) aid economic reading and reduce spurious sign inversions.

In policy and corporate contexts, the requirement is also procedural: a reproducible audit trail from data ingestion to publication; model version management; tracking of baselines adopted for IG; settings for SHAP; and bootstrap designs, with precise economic justifications. This documentation enables independent verification and shifts focus from the ``best model on average'' to the ``model adequate for the regime and robust to revision,'' a principle closer to the professional use of nowcasts.

The disciplined adoption of feature importance explainability and bootstrapping techniques renders ML models compatible with the responsibility constraints typical of business and economics: local explanations that anchor every update to identifiable drivers, global maps that maintain narrative consistency across regimes, and replicable intervals that document information quality at the moment of decision. Together, these elements transform a predictive output into an instrument effectively usable in environments where traceability, reproducibility, and information risk management are central.

\section[Governance and operational implications for business, markets, and policy]{Governance and operational implications for business, markets, and policy}
\label{sec:implications}

Nowcasting becomes a genuine decision-support tool only when it delivers timely, interpretable signals accompanied by robust uncertainty measures. These metrics must be integrated into public authorities' and private operators' workflows in a disciplined manner. In the presence of reporting lags, asynchronous releases, and informational ragged edge, a framework combining accuracy, strict traceability, and temporally coherent intervals reduces the information gap and mitigates the risk of overreacting to noisy signals.

Two pillars are indispensable. The first is structured estimate explainability, with local explanations for the latest update and global readings over the recent window. The second is uncertainty quantification based on time-aware procedures, such as block bootstrapping, rolling-window coverage diagnostics, and calibration checks. Absent these elements, a point forecast cannot be considered decision-grade.

\subsection{Policy use: central banks, ministries, and statistical institutes}
In the context of economic policy authorities, nowcasting informs decisions with medium- to short-term horizons but long-term consequences. A nowcasting system is sound only if it produces a minimum reporting set for each quarter: a \textbf{point estimate}, a \textbf{calibrated prediction interval}, and a \textbf{synthetic contribution map} by information blocks, such as domestic demand (consumption, investment, inventories), external demand (exports/imports, trade balance), labor market (employment, hours, claims), prices/inflation (consumer price index,  producer price indices), and financial conditions (rates, spreads, credit) \citep{PetropoulosMakridakis2020}. For high-impact decisions, reliance on thresholds defined relative to uncertainty bands is essential to reduce the probability of actions triggered by statistically fragile fluctuations.

Data asset management requires explicit snapshot management rules. In the absence of complete and homogeneous real-time archives, it is essential to freeze the statistical snapshot at predefined dates, archive versions, and prevent untracked retroactive modifications. That enables ex-post comparisons and independent audits of policy choices. The same discipline applies to updates between official releases: the treatment of ragged edge and high-frequency indicators must be documented, specifying how the nowcast is updated when partial new information arrives and which variables drive revisions.

Fallback protocols are an integral part of operational design. In phases where signal quality deteriorates---for instance, due to anomalous interval widening, loss of calibration, or systematic inconsistencies in drivers---the system must always provide a comparison with simple benchmarks (linear regularized models, RW with or without drift) \citep{MakridakisEtAl2020_M4,MakridakisSpiliotisAssimakopoulos2022}. These references anchor internal and external communication, clarifying the marginal value added by the ML framework over parsimonious solutions. In the presence of evident shocks or breaks, predefined protocols should be in place: scenarios with ``light'' re-estimation (on a recent window, assuming prudence), full re-estimation (adaptation to the new structure), and stress tests on critical information blocks. Each scenario must produce consistent estimates, intervals, and explanations, ensuring the final choice does not depend on the random outcome of a single model.

\subsection{Market and corporate use: financial intermediaries and firms}
For banks, asset managers, insurers, and large corporations, nowcasting typically fits into shorter decision chains, often embedded in daily or weekly risk dashboards. In this environment, model value lies in its ability to strictly update the business cycle assessment and link macro signals to explicit action rules: rebalancing only beyond specific probabilistic thresholds, sectoral adjustments if real-economic and financial drivers converge on a regime change, or revisiting internal stress scenarios when intervals and drivers signal a persistent increase in uncertainty.

Model risk management requires monitoring that distinguishes between model and data stability. Regarding model stability, it is essential to evaluate the temporal coherence of attributions produced by explainability tools: sudden or unjustified variations in feature contributions, given identical available information, are signs of estimation instability or excessive sensitivity to minor perturbations. Assessment must also consider the capacity of predictive intervals to maintain adequate coverage over rolling windows, a necessary condition for continuous operational use. On the data side, anomalies stemming from reporting issues (e.g., outliers, revisions, definition changes) must be rigorously distinguished from genuine forecast dynamics; diagnostics that conflate the two planes compromise both internal control solidity and the quality of communication to decision committees and supervisory functions.

It is also helpful to differentiate the informational perimeter by purpose: a readable ``core'' set suitable for communication with decision-making bodies, and an extended set to maximize accuracy in quantitative analysis units. In both cases, importance measures derived from XAI must be accompanied by uncertainty indications (e.g., resampling intervals) and not presented as rigid rankings. Time-series validation must be systematic: \textit{walk-forward} and \textit{backtesting} must reflect realistic temporal sequences; where historical vintages are available, financial institutions can go further, assessing robustness in true real-time \citep{HyndmanAthanasopoulos2021,Tashman2000}. In every exercise, out-of-sample reporting should be paired with deviation from parsimonious benchmarks to prevent interpretation from focusing solely on levels rather than informational gains.

\subsection{Governance, transparency, and deliverables}
A nowcasting system aiming for recurrent use in institutional or market contexts requires an explicit governance design. Separation of duties between data ingestion, modeling, validation, and publication reduces operational error risk and limits internal conflicts of interest. Version management rules must ensure that every modification to predictors, transformations, hyperparameters, or bootstrap procedures results in a new, traceable version, allowing historical comparisons only under identical configurations. Clear agreements on data update timing---both for official releases and high-frequency indicators---directly impact the perceived reliability of the system, especially when the nowcast enters regulatory processes or reporting obligations.

Transparency and communication require a minimum set of recurrent deliverables:
\begin{itemize}[leftmargin=*]
    \item the nowcast series with intervals and turning points;
    \item a waterfall decomposition of contributions for the latest update;
    \item an economic consistency table comparing signs and magnitudes of drivers with theoretical expectations for variable classes;
    \item a methodological appendix containing data sources and transformations, update rules, model specifications, design choices for explainability techniques (e.g., baselines for IG, thresholds for VIP), and essential details of the block bootstrap design.
\end{itemize}
A complete audit log (including data sources, release calendars, model versions, and calculation configurations) ensures traceability and is often required by regulators and internal control functions.

Signal quality is evaluated not only by average error but also by dynamic coherence. Near turning points, excessively narrow intervals, or flat importance measures indicate misspecification or informational leakage; symmetrically, arbitrary driver oscillations unsupported by macroeconomic fundamentals signal data or pipeline issues. A supervisory dashboard should therefore systematically include quantitative robustness indicators:
\begin{enumerate}[label=\roman*), leftmargin=*, itemsep=0pt]
    \item \textbf{Effective coverage} of intervals over rolling windows;
    \item\textbf{Importance stability}, summarized by a robust measure of central tendency (e.g., median importance) and a dispersion metric (e.g., interquartile range or bootstrap‑based confidence bands);
    \item \textbf{Distance from benchmarks}, via relative forecast errors, such as Root Mean Square Forecast Error (RMSFE) and Mean Absolute Forecast Error (MAFE);
    \item \textbf{Instability signals}, identified by unjustified changes in importance profiles between consecutive snapshots.
\end{enumerate}

Maintaining parsimonious, autoregressive, or naïve baselines is not merely a comparison exercise but a governance element: it defines the minimum thresholds that an explainable nowcasting system equipped with intervals must surpass to be considered genuinely suitable for operational use.

These considerations motivate a disciplined, end‑to‑end workflow for decision‑grade nowcasting systems, which we outline in the next section.

\section[An end-to-end workflow for decision-grade nowcasting systems]{An end-to-end workflow for decision-grade nowcasting systems}
\label{sec:workflow}

\subsection{Operational architecture}
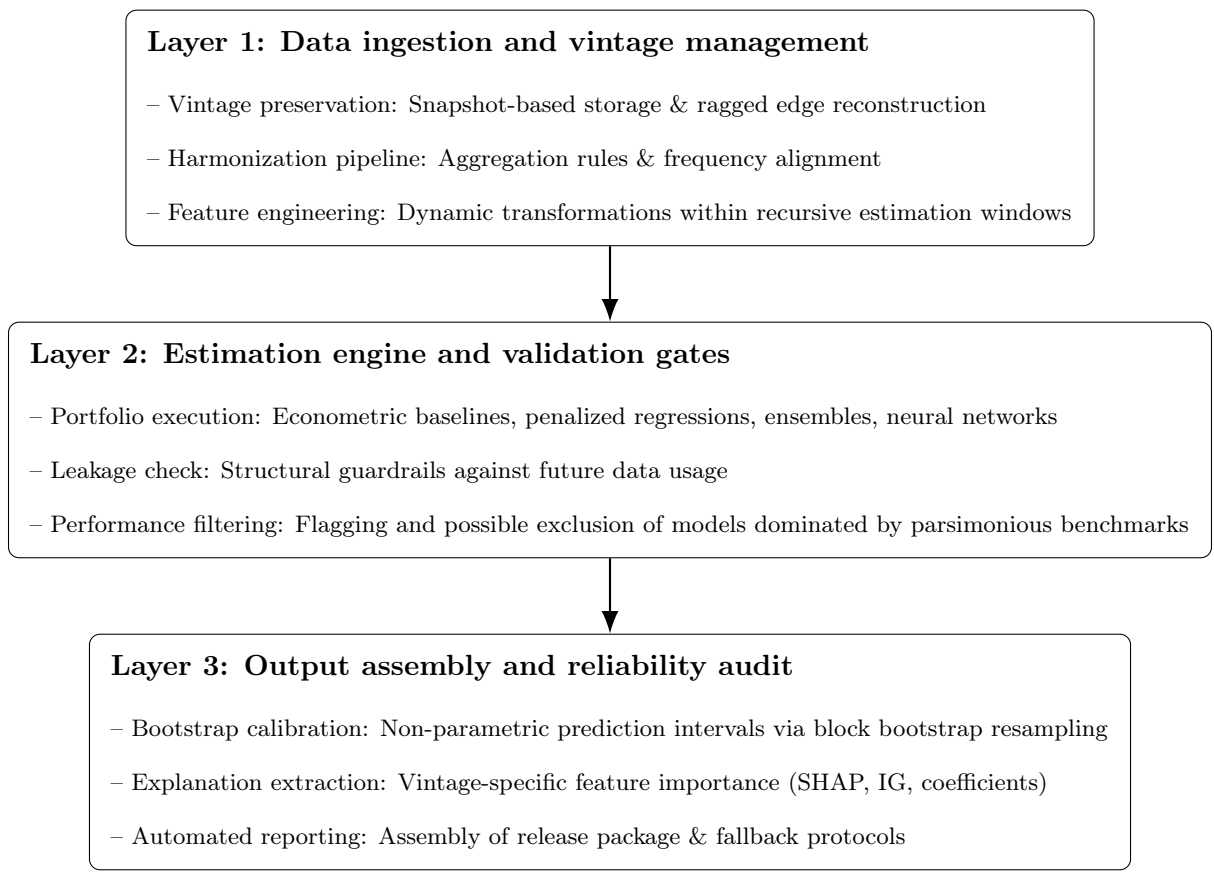
\begin{figure}[t]
    \centering
    \begin{tikzpicture}[
        box/.style = {
            rectangle,
            rounded corners,
            draw,
            align = left,
            minimum width = 12cm,
            inner sep = 8pt
        },
        arrow/.style = {
            -{Latex[length=3mm]},
            thick
        },
        node distance = 1.0cm
    ]

    \node[box] (data) {
        \textbf{Layer 1: Data ingestion and vintage management}\\[0.5ex]
        \footnotesize -- Vintage preservation: Snapshot-based storage \& ragged edge reconstruction\\
        \footnotesize -- Harmonization pipeline: Aggregation rules \& frequency alignment\\
        \footnotesize -- Feature engineering: Dynamic transformations within recursive estimation windows
    };

    \node[box, below=of data] (models) {
        \textbf{Layer 2: Estimation engine and validation gates}\\[0.5ex]
        \footnotesize -- Portfolio execution: Econometric baselines, penalized regressions, ensembles, neural networks\\
        \footnotesize -- Leakage check: Structural guardrails against future data usage\\
        \footnotesize -- Performance filtering: Flagging and possible exclusion of models dominated by parsimonious benchmarks
    };

    \node[box, below=of models] (xai) {
        \textbf{Layer 3: Output assembly and reliability audit}\\[0.5ex]
        \footnotesize -- Bootstrap calibration: Non-parametric prediction intervals via block bootstrap resampling\\
        \footnotesize -- Explanation extraction: Vintage-specific feature importance (SHAP, IG, coefficients)\\
        \footnotesize -- Automated reporting: Assembly of release package \& fallback protocols
    };

    \draw[arrow] (data.south) -- (models.north);
    \draw[arrow] (models.south) -- (xai.north);

    \end{tikzpicture}
    \caption{Conceptual workflow for decision-grade nowcasting.}
    \label{fig:workflow}
\end{figure}

This section translates the governance requirements discussed in Section~\ref{sec:implications} into an operational engineering pipeline. The workflow is designed as a reference architecture to ensure reproducibility and to enforce quality controls on data handling and model validation. Figure~\ref{fig:workflow} outlines the three logical layers: data ingestion, estimation engine, and output assembly.

\subsection{Layer 1: Data ingestion and vintage management}
The foundational layer strictly separates information availability from model consumption to prevent look-ahead bias.
\begin{itemize}[leftmargin=*]
    \item \textbf{Vintage Preservation:} The system should rely on a snapshot-based storage mechanism. Every run retrieves data as it existed on the specific reference date, reconstructing the exact ragged edge structure available to a real-time observer.
    \item \textbf{Harmonization Pipeline:} High-frequency flows and monthly series are aggregated to the target frequency using deterministic rules (e.g., temporal averaging, end-of-period slicing) defined ex-ante to ensure consistency across updates.
    \item \textbf{Feature Engineering:} Transformations are applied dynamically within each \textit{recursive estimation window}. This prevents information leakage from the full sample into the training set. The layer outputs a frozen feature set aligned with the target variable at each forecast origin, ready for ingestion by the estimation engine.
\end{itemize}

\subsection{Layer 2: Estimation engine and validation gates}
This layer executes the model portfolio in parallel, enforcing time-aware validation protocols.
\begin{itemize}[leftmargin=*]
    \item \textbf{Portfolio Execution:} The system trains the defined set of candidate models (econometric baselines, penalized regressions, tree ensembles, neural networks) on expanding or rolling windows that respect the temporal order of observations.
    \item \textbf{Leakage Check:} A distinct validation step verifies that no predictor contains information timestamped after the forecast origin, acting as a structural guardrail against inadvertent future data usage.
    \item \textbf{Performance Filtering:} Models are evaluated against parsimonious benchmarks (e.g., Random Walk or AR). Specifications that are consistently dominated in terms of RMSFE or MAFE over the validation window are flagged for review and, if necessary, excluded from the final ensemble to maintain predictive quality.
\end{itemize}

\subsection{Layer 3: Output assembly and reliability audit}
The final layer integrates raw predictions with interpretability and uncertainty modules to produce a decision-grade signal.
\begin{itemize}[leftmargin=*]
    \item \textbf{Bootstrap Calibration:} For every retained model, the block bootstrap module generates $B$ resampled trajectories. This allows for the construction of a $(1-\alpha)\%$ prediction interval where the width reflects the volatility of the current regime rather than a static error assumption.
    \item \textbf{Explanation Extraction:} Feature importance scores (via SHAP, IG, or coefficients/loadings) are computed relative to the specific vintage, attributing the nowcast level to specific economic indicator blocks.
    \item \textbf{Automated Reporting:} The system assembles three components into a final release package: the point forecast, the calibrated prediction interval, and the driver decomposition. If the uncertainty band exceeds a predefined tolerance level, indicating lack of convergence or extreme instability, the workflow triggers fallback protocols (e.g., reverting to parsimonious benchmarks and flagging the ML signal as low confidence).
\end{itemize}

\section[Research agenda and gaps]{Research agenda and gaps}
\label{sec:agenda}

Relative to existing surveys on macroeconomic nowcasting and ML forecasting, the contribution of this article is to connect model choice, feature importance explainability, and block-bootstrap uncertainty quantification within a governance-oriented, end-to-end workflow for decision-grade deployment.

Within this framework, several questions remain open for further research. Patterns emerging from the empirical applications in Table~\ref{tab:ml_nowcasting_lit} and from the cross-family framework in Table~\ref{tab:xai_uncertainty_framework} reveal substantial heterogeneity in model choice, explainability tools, and bootstrap designs, which directly motivates several of the research directions discussed below.

Applications of macroeconomic and macro-financial nowcasting require an operational balance between accuracy, transparency, and uncertainty management. The pressing questions concern not which model ``prevails'' on average, but when, why, and under what regimes specific classes prove more reliable; how to make predictive intervals comparable in the presence of marked temporal dependencies; and how to obtain stable explanations without introducing leakage or violating economic identities.

A primary avenue for further inquiry involves multi-model evaluation across differentiated regimes \citep{MedeirosEtAl2021,CoulombeEtAl2022}. Evidence suggests that accuracy, interval coverage, and attribution stability vary significantly with changing macroeconomic conditions. Time-aware replications and cross-checks on sudden shocks can clarify the relative sensitivity of penalized models, factor specifications, ensembles, and neural networks, offering robust criteria to determine which families maintain consistent performance out-of-regime \citep{InoueRossi2012,ClarkMcCracken2009}. Theoretically, PAC-Bayes approaches and sequential aggregation establish oracle-type inequalities for forecasting weakly dependent series, providing proof-of-concept guarantees for model selection and combination in dependent contexts \citep{alq_wint_fast,alquier2012prediction,BanerjeeRaoHonnappa2021}.

A second research directive concerns the design of the block bootstrap. The joint selection of block length and structure remains insufficiently codified in the literature, particularly in the presence of ragged edge, latent components, or asynchronous publication calendars \citep{PolitisRomano1994,Li2021}. Comparative studies based on equivalent information windows could help isolate configurations that guarantee intervals consistent with economic dynamics without introducing excessive resampling variability.

A third requirement emerges from the comparability of explainability measures across modeling families \citep{LundbergEtAl2020,SundararajanTalyYan2017}. Coefficients and Variable Importance in Projection for penalized and dimension-reduction models, impurity- or gain-based importances for ensembles, and \textit{Integrated Gradients} for neural networks convey information that is not always immediately aggregable. A common metric for stability, sign coherence, and economic traceability would facilitate the use of explanations over rolling horizons and enable structured comparisons between heterogeneous approaches.

A further strand concerns nowcast selection and combination. One strategy involves applying the Model Confidence Set (MCS) \citep{HansenLundeNason2011}, estimated via block bootstrapping on out-of-sample losses, to define a subset of models with equal predictive ability and combine their nowcasts through averaging (simple, weighted inversely proportional to cumulative errors, or exponentially weighted based on cumulative losses). Evaluation is performed via \textit{walk-forward} backtesting over the entire out-of-sample period, reconstructing the evolution of weights over time and comparing the trajectories of combined models with those of single models selected by the MCS. Future developments could explore rolling versions of the MCS and more regularized weighting rules to contain weight turnover and avoid myopic reactions to recent shocks. In current empirical applications, the MCS is generally estimated once over a sufficiently long out-of-sample horizon, producing a ``global'' set of non-discarded models; extending the procedure to rolling windows is a natural adaptation for contexts characterized by marked instability.

Transferability to small, open economies, particularly in geographically less-explored areas such as Southeast Asia, remains a relevant research theme. In contexts exposed to frequent global shocks and characterized by relatively short databases, a disciplined framework with targeted interventions on explainability and bootstrapping may offer advantages over standard solutions designed for large economies with long time series. However, tests on systems with different informational structures are necessary to disentangle regime-specific elements from those that are genuinely generalizable.

These lines intersect with a cross-cutting issue: the standardization of experimental reporting. Replicable protocols for temporal splitting, update rules between releases, hyperparameter specifications, audit criteria for feature importance explainability, and bootstrap settings would improve comparability between studies and limit artifactual components deriving from non-standardized pre-processing, configuration, and validation choices. The integration of quantitative indicators (adequate coverage across rolling windows, stability of importance via robust statistics, relative distance from benchmarks, and signals of temporal instability in predictor-target relationships) provides a richer framework for assessing the system's operational reliability.

Empirically, the growing availability of high-frequency indicators opens avenues for testing the timeliness and resilience of diagnoses, verifying whether the inclusion of denser flows increases sensitivity to turning points without compromising attribution stability or interval coherence. The same logic guides the exploration of new modeling architectures: relevance lies not only in nowcasting accuracy but also in the capacity to maintain the explanatory coherence of results and to quantify meaningful predictive uncertainty under realistic data-release conditions.

\section*{Acknowledgments}

I am grateful to Rosaria Romano (University of Naples Federico II), Jamus Jerome Lim (ESSEC Business School Asia-Pacific), and Pierre Alquier (ESSEC Business School Asia-Pacific) for their guidance throughout my doctoral studies and for valuable discussions that helped shape the broader research agenda from which this paper draws. The analysis and any remaining shortcomings are entirely my own.

\bibliography{refs}

\end{document}